\documentclass[aps,twocolumn,showpacs,superscriptaddress]{revtex4}
\usepackage[dvips]{graphicx}
\usepackage{amsmath}
\usepackage{amssymb}
\usepackage{dcolumn}
\begin{document}
\title{Velocity distributions in dissipative granular gases}
\author{J.\ S.\ van Zon}
\author{F.\ C.\ MacKintosh}
\affiliation{Division of Physics and
Astronomy, Vrije Universiteit, 1081 HV Amsterdam, The Netherlands}
\affiliation{Kavli Institute for Theoretical Physics, University of
California, Santa Barbara, CA 93106}
\date{\today}
\begin{abstract}
Motivated by recent experiments reporting non-Gaussian velocity distributions in driven dilute granular materials, we study by numerical simulation the properties of 2D inelastic gases. We find theoretically that the form of the observed velocity distribution is governed primarily by the coefficient of restitution $\eta$ and $q=N_H/N_C$, the ratio between the average number of heatings and the average number of collisions in the gas. The differences in distributions we find between uniform and boundary heating can then be understood as different limits of $q$, for $q \gg 1$ and $q \lesssim 1$ respectively.
\end{abstract}
 \maketitle

Granular materials consisting of macroscopic particles or grains can exhibit behavior reminiscent of conventional phases of matter. Sand, for instance, can flow like a liquid under some conditions. Dilute granular systems, or {\em gases}, have been extensively studied both experimentally and theoretically, in large part as simple model systems exhibiting nonequilibrium and dissipative behavior. These systems are intrinsically dissipative and out of equilibrium, even though it is tempting to apply such equilibrium notions as temperature. Since the collisions in such a gas are inelastic, a gas-like steady state is achieved only with a constant \emph{drive}, or input of energy.  Otherwise, all motion ceases after only a finite time \cite{mcnamara,kadanoff1}. In principle, it is possible to drive the system uniformly throughout the container, with every particle in contact with a heat source all the time ({\em uniform heating}). This has been done in simulations \cite{mackintosh,moon} and is assumed in analytic theories \cite{ernst}. In experiments, however, one usually drives a granular gas by shaking or vibrating the walls of the container. Such {\em boundary heating} means that the energy is inserted in a spatially inhomogeneous way \cite{luding,rouyer,kudrolli}. As a consequence, the gas will develop a gradient in density and mean kinetic energy \cite{kadanoff2}. Even for uniform heating, however, significant deviations from equilibrium gases, e.g., in density correlations, are observed \cite{mackintosh}. 

One of the most fundamental aspects of molecular gases is the Maxwell-Boltzmann velocity distribution \cite{ReifOrOtherText}. A very general and striking feature of driven dissipative gases, however, is the apparently strong deviation from this classical behavior. Such velocity distributions have been the subject of numerous recent experimental studies \cite{luding,rouyer,losert,kudrolli}, and characteristic velocity distributions of the form $P(v)=C \exp[-\beta(v/\sigma)^{\alpha}]$ have been observed, where $\sigma=\langle v^2\rangle ^{\frac{1}{2}}$ is often called the granular temperature, in analogy with equilibrium gases. Experiments of Rouyer and Menon \cite{rouyer}, in particular, have suggested that a universal exponent of $\alpha=1.5$ occurs over a wide range of experimental parameters. This observation was particularly intriguing, as Van Noije and Ernst \cite{ernst} predicted a high-velocity tail with an exponent 1.5 using kinetic theory. Here, we demonstrate instead that a family of distributions showing apparent exponents covering a wide range of values $\alpha<2$ is expected, depending on both material and experimental conditions. Furthermore, we show that the velocity distribution is governed primarily by the relative importance of collisions to heating, \emph{i.e.}, the way in which energy flows through the system of particles. Specifically, we introduce a new parameter $q=N_H/N_C$, which measures the ratio between numbers of heating events and collisions experienced by a typical particle. These theoretical observations can explain both the observed non-Gaussian behavior, as well as the ambiguities in the experimental and theoretical literature on dissipative gases to date. We also demonstrate that the behavior of the velocity distributions can be captured quantitatively by a simple model that takes only $\eta$ and $q$ into account, with no spatial degrees of freedom.

Even though the distribution of velocities is such a basic property of gases, results for velocity distributions in dissipative systems have remained ambiguous. Both in simulation and experiment, different setups and driving mechanisms usually give different behavior of the velocity distribution. For a setup where particles on a horizontal plate were driven in the vertical direction, Olafsen and Urbach \cite{olafsen} found a crossover from exponential to Gaussian distributions as the amplitude of the driving was increased. The result of Rouyer and Menon \cite{rouyer} was obtained for a different configuration where particles were confined between two vertical plates and driven in the vertical direction. Blair and Kudrolli \cite{kudrolli} use a different setup where particles move along an inclined plane. Friction with the plane during collisions reduces the effective coefficient of restitution to $\eta\approx0.5$, in contrast to $\eta=0.93$ in the setup of Rouyer and Menon. They find the distribution with exponent $\alpha=1.5$ only in the very dilute case. Otherwise, the distributions deviate strongly from both Gaussian and the distribution obtained by Rouyer and Menon. For the denser case Blair and Kudrolli find a distribution that has one exponent for low velocities and crosses over to another for high velocities. This crossover is also observed in  simulation by Moon, et al. \cite{moon}.

Thus, it remains unclear what the analogue (if any) of the Maxwell-Boltzmann distribution is for dissipative gases, let alone what the origin of the non-Gaussian behavior is. Puglisi, et al. \cite{puglisi} have suggested that the deviations are caused by the spatial correlations in the gas. They propose that for a region of uniform density the velocity distribution of the gas actually is Gaussian, with a density dependent width. The spatial correlations cause density fluctuations and they claim the non-Gaussian distributions arise as an average over the velocity distributions over these regions of different density. Experiments \cite{rouyer, olafsen} show that this is not the case. Even within windows of uniform density the velocity distributions remain non-Gaussian. Our results below confirm this.

We study velocity distributions using an event-driven simulation of $N$ particles of radius $r$ moving in a two-dimensional box. Particles gain energy by heating and lose energy through inelastic collisions. When two particles $i$ and $j$ collide their final velocities depend on their initial velocities in the following way: 
\begin{equation}
\mathbf{v}_{i}^{\prime}=\mathbf{v}_{i}-\frac{1+\eta}{2}(\mathbf{v}_{i}
\cdot \mathbf{\hat{r}}_{ij}-\mathbf{v}_{j} \cdot
\mathbf{\hat{r}}_{ij})\mathbf{\hat{r}}_{ij},
\label{eqn:inelas}
\end{equation}
where $\eta$ is the coefficient of restitution and $\mathbf{\hat{r}}_{ij}$ is the unit vector connecting the centers of particles $i$ and $j$.

When heating uniformly, we add every time step $\Delta t$ a random contribution to the velocity of all the particles and we use periodic boundary conditions to simulate bulk behavior. Note that this is significantly different from the spatially homogeneous heating used in experiments \cite{olafsen}, as here in uniform heating the driving is uncorrelated in space and time. When heating through the boundary, particles are confined in a box with radius $R=1$. The particles receive a kick upon collision with the boundary. We assume that the collision between particles and the boundary is elastic and we add a random contribution to the velocity in the direction perpendicular to the boundary. For more details, we refer to \cite{jeroen}. We allow the system to reach steady state before taking data. For both uniform and boundary heating, data is taken periodically every $\Delta t$.

\begin{figure}[ht] \centering
\includegraphics[width=7cm]{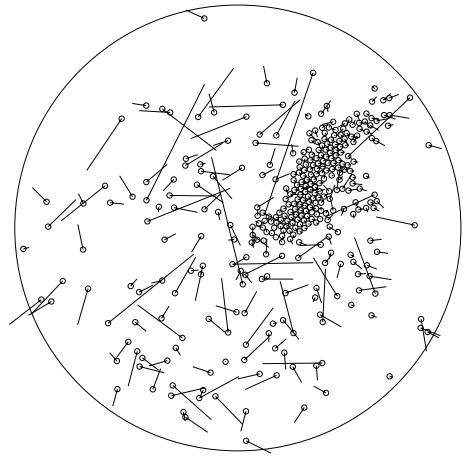} \caption{Snapshot of a clustered state state for $N=350$, $\phi=0.05$ and $\eta=0.6$. Particles are indicated by circle, while the lines show the direction and magnitude of the velocity.}
\label{fig:cluster} \end{figure}

One of the first striking differences between uniform heating and boundary heating is clustering. When heating through the boundary, a stable liquid-like cluster surrounded by a hot gaseous state will form for low coefficients of restitution $\eta$ or high area fraction $\phi$. This occurs as particles are compressed in the center of the box by particles moving in from the boundary. The increase of density leads to collapse and a stable cluster is formed. A typical example is shown in Fig.~\ref{fig:cluster}. These clusters do not occur in our simulations with uniform heating. This is because particles are heated all the time, which prevents the collapse to a cluster.

\begin{figure}[hb] \centering
\includegraphics[width=\columnwidth]{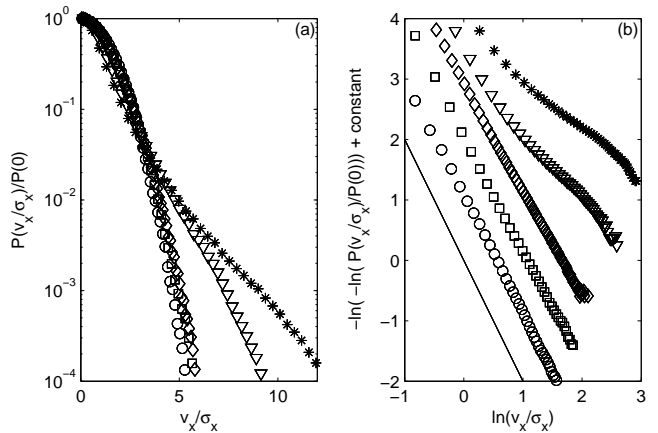} \caption{Velocity distributions for $N=350$ and $\phi=0.02$. Shown are both results for uniform heating with $\eta=0.8$($\circ$), $\eta=0.1$($\square$) and results for boundary heating with $\eta=0.9$($\diamondsuit$), $\eta=0.6$($\triangledown$) and $\eta=0.4$($*$). (a) $P(v_x/\sigma_x)$ (b) $-\ln\{-\ln[P(v_x/\sigma_x)]\}$ versus $\ln(v_x/\sigma_x)$. A Gaussian is shown as a solid line.}
\label{fig:uniform_boundary} \end{figure}

Velocity distributions obtained for uniform and boundary heating are shown in Fig.~\ref{fig:uniform_boundary}. When heating through the boundary, the gas develops a gradient in area fraction and granular temperature. In this case, we record the velocity distribution in a ring of radius $0.4<r\leq0.6$ around the center of the box, where the granular temperature is approximately constant. The velocity distributions in the different rings varies only slightly. For uniform heating, we collect data everywhere in the box.

Fig.~\ref{fig:uniform_boundary} shows clear qualitative differences between uniform and boundary heating. For uniform heating the velocity distribution is close to Gaussian for a large range of $\eta$. In contrast, for boundary heating the distribution is only Gaussian in the nearly elastic case $\eta=0.9$. As the coefficient of restitution is lowered, a crossover develops where the exponent changes from $\alpha_1 \approx 2$ to a lower value $\alpha_2$. The exponent $\alpha_2$ becomes smaller as dissipation is increased (for smaller $\eta$ or higher $\phi$) and we find any value in the range $0.7\lesssim\alpha_2<2$. As $\alpha_2$ decreases it becomes increasingly difficult to describe the high velocity tail with a single exponent $\alpha_2$. This crossover in exponent is observed in boundary heating for all values of $\phi$ and $N$.

\begin{figure}[b] \centering 
\includegraphics[width=7.5cm]{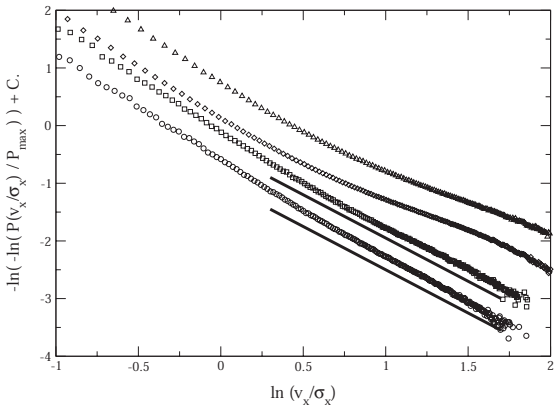} \caption{(a) $-\ln\{-\ln[P(v_x/\sigma_x)]\}$ versus $\ln(
v_x/\sigma_x)$ for $N=350$, $\phi=0.05$ and $\eta=0.9$ ($\circ$), $N=500$, $\phi=0.05$ and $\eta=0.9$ ($\square$), $N=350$, $\phi=0.05$ and $\eta=0.8$ ($\diamond$), $N=350$, $\phi=0.25$ and
$\eta=0.9$ ($\triangle$). The solid lines correspond to the fit as made by Rouyer and Menon and has an exponent $\alpha=1.52$. The range of the solid lines corresponds to half the range used by Rouyer and Menon in their fit, but contains about 80$\%$ of their data points.} \label{fig:rouyer}
\end{figure}
 
For a certain range of parameters we also find high-velocity tails with exponent $\alpha_2=1.5$. For their experiments Rouyer and Menon used $N$ particles with $\eta\approx0.9$, where $100<N<500$ and $0.05<\phi<0.25$ \cite{rouyer}. In Fig.~\ref{fig:rouyer} we plotted the velocity distribution for $\eta=0.9$, $\phi=0.05$ and several values of $N$. We also show the fit with $\alpha=1.52$ as made in \cite{rouyer}. This line clearly coincides with the high-velocity tail of the velocity distribution found by the present simulation. This suggests that instead of a universal distribution with $\alpha=1.5$, they might have observed a part of a more complex velocity distribution, with more than one apparent exponent. 

The main difference between uniform and boundary heating is that in the first case heating takes place homogeneously throughout the box, whereas in the latter case energy is injected inhomogeneously at the boundaries. This is not the direct cause for the difference in velocity distributions. When heating homogeneously, one can go from a Gaussian distribution to one with a crossover by increasing the time between heatings \cite{jeroen} allowing the average number of heatings per unit time to become smaller than the average number of collisions. The reverse is also true. When heating inhomogeneously through a boundary, one finds Gaussian distributions in very dilute systems, when the particles on average collide more often with the boundary than with other particles. This suggests that in our system the shape of the velocity distribution is not a function of all parameters $\eta$, $\phi$, $N$ and the details of the energy injection, but only of $\eta$ and $q=N_H/N_C$, the ratio between the average number of heatings $N_H$ and the average number of collisions $N_C$.

\begin{figure}[ht] \centering 
\includegraphics[width=7.5cm]{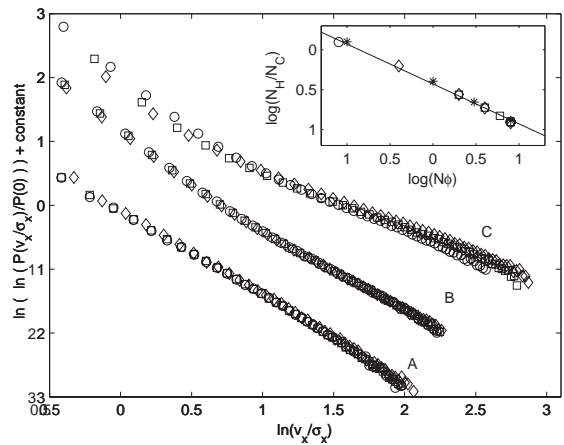} \caption{ Velocity distributions for different values of the heating-dissipation rate $q$. Distributions with the same $q$ are shown on top of each other. (A) $q=1.3$ and we show $N=100$ and $\phi=1\cdot10^{-3}$($\circ$), $N=200$ and $\phi=5\cdot10^{-4}$($\square$), $N=800$ and $\phi=1.25\cdot10^{-4}$($\diamondsuit$). (B) $q=0.13$ and we show $N=100$ and $\phi=0.08$($\circ$), $N=200$ and $\phi=0.04$($\square$), $N=400$ and $\phi=0.02$($\diamondsuit$). (C) $q=0.013$ and we show $N=100$ and $\phi=0.4$($\circ$), $N=200$ and $\phi=0.2$($\square$), $N=400$ and $\phi=0.1$($\diamondsuit$). Inset: Heating-dissipation ratio $q$ for $N=800$($\circ$), $N=400$($\square$), $N=200$($\diamondsuit$) and $N=100$($*$) for several values of $\phi$. The line is a fit of the form $(N \phi)^{1/2}$.}  \label{fig:diff_f}
\end{figure}

This idea can easily be tested in boundary heating. When increasing the number of particles $N$ or the area fraction $\phi$, the average number of collisions increases. One can show in a mean field approximation that $q \sim (N \phi)^{-1/2}$. The average distance a particle travels between collisions is given by $l_{coll} \sim 1/\phi$. For a box of area $A$ the average distance between boundaries is given by $l_{heat} \sim A^{1/2} \sim (N/\phi)^{1/2}$. Finally, we know $N_H/N_C \sim l_{coll}/l_{heat}$. Our simulation obeys this approximation very well. In Fig.~\ref{fig:diff_f} we show velocity distributions for $\eta=0.8$ and different combinations of $N$ and $\phi$. We measure the heating-dissipation ratio $q$ in the simulation and show velocity distributions with the same $q$ on top of each other. For $q=1.3$ and $q=0.13$ we find excellent collapse for different $N$ and $\phi$, even when we scale the system by a factor $8$. For $q=0.013$, where spatial correlations become very strong, we still find reasonable collapse. As we increase $q$ we observe the usual pattern, where a crossover appears in a distribution that was initially close to a Gaussian.

\begin{figure}[ht] \centering 
\includegraphics[width=7.5cm]{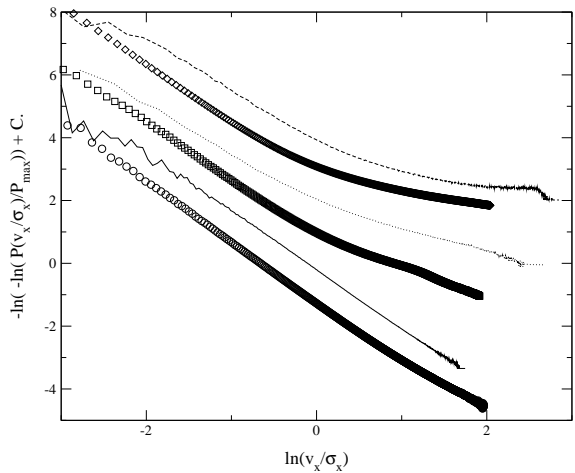} \caption{ $-\ln\{-\ln[P(v_x/\sigma_x)]\}$ versus $\ln(v_x/\sigma_x)$. The symbols shown are velocity distributions acquired by simulation for $q=120$ ($\circ$), $0.08$ ($\square$), $0.012$ ($\diamond$). The lines show the velocity distributions found in the model for the same values of $q$ (solid, dotted, dashed).}  \label{fig:fcompare}
\end{figure}

It has been speculated that the non-Gaussian distributions are caused by spatial correlations in the gas \cite{puglisi}. However, we can qualitatively reproduce the different distributions we observe in simulation with a simple model (based on Ref. \cite{ulam}, but including dissipation) of a two-dimensional inelastic gas of $N$ particles without spatial degrees of freedom. In this model, every time step $C$ pairs of particles are selected at random and collide using equation \ref{eqn:inelas} with a random impact parameter $-2r<b<2r$, where $r$ is the radius of the particles. At the same time $H$ particles are selected at random to be heated by adding a random velocity. This gives us a heating-dissipation ratio of $q=H/C$. In Fig.~\ref{fig:fcompare} we compare velocity distributions from the model and simulations for different values of $q$. We find good qualitative agreement between simulation and model, even though it has no spatial degrees of freedom. This implies that spatial correlations play a minor role, if any, in $P(v)$.

Due to the idealized nature of our system, it is not possible to do a direct comparison between our simulation and experiments. Yet, some experiments seem to show a similar behavior, accounting for the parameter $q$ that we introduced here. In the experiments of Ref. \cite{olafsen} velocity distributions go from non-Gaussian to Gaussian when a rough plate or a layer of heavy particles is used instead of a flat plate. In this case, energy is injected directly into the directions parallel to the plate, effectively increasing the number of in-plane heatings over collisions. Most convincing is the experiment by Blair and Kudrollli \cite{kudrolli}. Here the number of collisions is increased by adding more particles. As a result, their velocity distributions develop the same crossover we see both in our simulations and model.   

We studied behavior of the velocity distributions of granular gases as a function of $\phi$, the area fraction, and $\eta$, the coefficient of restitution. Specifically, we considered the effect of driving the gas by heating uniformly, as is assumed in theory and many prior simulations, and by heating through a boundary, as is done in most experiments. We find that there exists clear qualitative differences between the velocity distributions for uniform and boundary heating, which demonstrates that the form of the distributions is not simply a function of material parameters (\emph{e.g.}, $\eta$). Furthermore, we show that there is no evidence for a universal velocity distribution with a constant exponent $\alpha=1.5$. Instead, for boundary heating, we find that velocity distributions cross over from one exponent to another for the high-velocity tail. For this high-velocity tail we observe a wide range of exponents and we find $\alpha=1.5$ only for specific values of $\phi$ and $\eta$.

Instead, we demonstrate that the distribution of velocities for dissipative gases, while not universal in form, depends only on two parameters: the coefficient of restitution $\eta$ (a material parameter) and $q=N_H/N_C$, the average ratio of heatings and collisions in the gas (a function of experimental conditions). We find that velocity distributions range from Gaussian for $q \gg 1$, where heating dominates dissipation, to strongly non-Gaussian for $q \ll 1$, where the dynamics of the gas is dominated by the dissipative collisions between particles. Furthermore, a simple model of a driven, inelastic gas without spatial degrees of freedom reproduces the entire family of velocity distributions we find in simulation, as we vary $\eta$ and $q$. This means that the velocity distributions are non-Gaussian not because of spatial correlations. Rather, it is the cascade of energy from a few high-energy particles to the slow-moving bulk of the gas that is the key determinant of the non-Gaussian velocity distributions.

We thank L.P. Kadanoff, N. Menon, A. Kudrolli, D. Blair  and J. Bougie for useful conversations. This work is supported in part by the National Science Foundation under Grant PHY99-07949.

\end{document}